\documentstyle{article}

\makeatletter

\def\@aabuffer{}
\def\author #1{\expandafter\def\expandafter\@aabuffer\expandafter
{\@aabuffer \small\rm      #1\relax \par}}
\def\address#1{\expandafter\def\expandafter\@aabuffer\expandafter
{\@aabuffer \small\it #1\relax \par\vspace{1em}}}

\def\maketitle{
\begin{center}
   {\bf \@title \par}
   \vskip 2em                      
   \@aabuffer\relax
\end{center} \par
\gdef\@aabuffer{}
}

\def\abstracts#1{
\begin{center}
{\begin{minipage}{4.2truein}
                 \footnotesize
                 \parindent=0pt #1\par
                 \end{minipage}}\end{center}
                 \vskip 2em \par}

\fussy
\def\section{\@startsection {section}{1}{\z@}{-3.5ex plus -1ex minus
    -.2ex}{2.3ex plus .2ex}{\bf }}
\def\subsection{\@startsection{subsection}{2}{\z@}{-3.25ex plus -1ex minus
   -.2ex}{1.5ex plus .2ex}{\it }}

\def\@makefnmark{{$\!^{\@thefnmark}$}}

\renewenvironment{thebibliography}[1]
	{\begin{list}{\arabic{enumi}.}
	{\usecounter{enumi}\setlength{\parsep}{0pt}
	 \setlength{\itemsep}{0pt}
         \settowidth
	{\labelwidth}{#1.}\sloppy}}{\end{list}}

\newcounter{arabiclistc}

\def\@citex[#1]#2{\if@filesw\immediate\write\@auxout
	{\string\citation{#2}}\fi
\def\@citea{}\@cite{\@for\@citeb:=#2\do
	{\@citea\def\@citea{,}\@ifundefined
	{b@\@citeb}{{\bf ?}\@warning
	{Citation `\@citeb' on page \thepage \space undefined}}
	{\csname b@\@citeb\endcsname}}}{#1}}

\newif\if@cghi
\def\cite{\@cghitrue\@ifnextchar [{\@tempswatrue
	\@citex}{\@tempswafalse\@citex[]}}
\def\citelow{\@cghifalse\@ifnextchar [{\@tempswatrue
	\@citex}{\@tempswafalse\@citex[]}}
\def\@cite#1#2{{$\!^{#1}$\if@tempswa\typeout
	{IJCGA warning: optional citation argument
	ignored: `#2'} \fi}}

\def\baselinestretch{1.0}
\ifx\selectfont\undefined
\@normalsize\else\let\glb@currsize=\relax\selectfont
\fi

\ifx\selectfont\undefined
\def\@singlespacing{%
\def\baselinestretch{1}\ifx\@currsize\normalsize\@normalsize\else\@currsize\fi%
}
\else
\def\@singlespacing{\def\baselinestretch{1}\let\glb@currsize=\relax\selectfont}
\fi

\long\def\@makecaption#1#2{
   \vskip 10pt
   \setbox\@tempboxa\hbox{\footnotesize #1: #2}
   \ifdim \wd\@tempboxa >\hsize   
       \leftskip 0pt plus 1fil
       \rightskip 0pt plus -1fil
       \parfillskip 0pt plus 2fil
       \footnotesize #1: #2\par   
     \else                        
       \hbox to\hsize{\hfil\box\@tempboxa\hfil}
   \fi}

\def\eqnarray{\stepcounter{equation}\let\@currentlabel=\theequation
\global\@eqnswtrue
\global\@eqcnt\z@\tabskip\@centering\let\\=\@eqncr
$$\halign to \displaywidth\bgroup\@eqnsel\hskip\@centering
  $\displaystyle\tabskip\z@{##}$&\global\@eqcnt\@ne 
  \hfil$\displaystyle{\hbox{}##\hbox{}}$\hfil
  &\global\@eqcnt\tw@ $\displaystyle\tabskip\z@
  {##}$\hfil\tabskip\@centering&\llap{##}\tabskip\z@\cr}

\def\lefteqn#1{\hbox to 2em{$\displaystyle #1$\hss}}

\makeatother

\def\Journal#1#2#3#4{{#1} {\bf #2}, #3 (#4)}
\def\mbar#1{\kern 0.1em\overline{\kern -0.1em #1 \kern -0.1em} 
  \kern 0.1em}

\begin{document}

\title{ON SOME CLASS OF MULTIDIMENSIONAL NONLINEAR INTEGRABLE
SYSTEMS\footnote{Talk given at the II$^{\rm nd}$ International
Sakharov Conference on Physics, May, 20--24, 1996, Moscow, Russia.}}

\author{ A.V. RAZUMOV, M.V. SAVELIEV }

\address{Institute for High Energy Physics, Protvino,
Moscow region,\\ 142284, RUSSIA}

\maketitle\abstracts{On the base of Lie algebraic and differential geometry
methods, a wide class of multidimensional nonlinear integrable systems is 
obtained, and the integration scheme for such equations is proposed.}

{\bf 1.} In the report we give a Lie algebraic and differential geometry
derivation of a wide class of nonlinear integrable systems of partial
differential equations for the functions depending on an arbitrary number of 
variables, and construct, following the lines of Refs. 1--3,
their general solutions in a `holomorphically factorisable' form. The systems 
are generated by flat connections, constrained by the relevant grading 
condition, with values in an arbitrary reductive complex Lie algebra
${\cal G}$ endowed with a ${\bf Z}$--gradation. They describe a 
multidimensional version of Toda type fields coupled to matter fields, 
and, analogously to the two dimensional situation, with an appropriate 
In\"on\"u--Wigner contraction procedure, for our systems one can exclude back 
reaction of the matter fields on the Toda fields. 

For two dimensional case and the connection taking values in the
local part of a finite dimensional algebra ${\cal G}$, our equations
describe an (abelian and nonabelian) conformal Toda system and its
affine deformations for an affine ${\cal G}$, see Ref. 1 and
references therein, and also Ref. 2 for differential and algebraic
geometry background of such systems. For the connection with values
in higher grading subspaces of ${\cal G}$ one deals with systems
discussed in Refs. 3, 4. In higher dimensions our systems, under some
additional specialisations, contain as particular cases the
Cecotti--Vafa type equations\cite{CVa91} written there for a case of
the complexified orthogonal algebra, see also Ref. 6; and those of
Gervais--Matsuo\cite{GMa93} which represent some reduction of a
generalised WZNW model. Due to the lack of space we present here only
an announcement of the results which will be described in detail,
together with some remarkable examples elsewhere.

{\bf 2.} Let $M$ be the manifold ${\bf R}^{2A}$ with the standard coordinates 
$z^{\pm i}, 1\leq i\leq A$, or ${\bf C}^A$ with $z^{+i} =\mbar{z^{-i}}$; let 
$G$ be a reductive complex Lie group with the Lie algebra ${\cal G}$ endowed 
with a ${\bf Z}$--gradation, ${\cal G}=\oplus_{m\in{\bf Z}}{\cal G}_m$. 
Consider a flat connection $\omega =\sum_{i=1}^A (\omega_{-i} dz^{-i} + 
\omega_{+i} dz^{+i})$ on the trivial principal fiber bundle $M\times G \to M$, 
and impose on it the grading condition that the components $\omega_{\pm i}$ 
take values in ${\cal G}_0\oplus
{\cal N}_{\pm i}$, where ${\cal N}_{\pm _i}=\bigoplus_{1\leq m\leq l_{\pm i}} 
{\cal G}_{\pm m}$ with some positive integers $l_{\pm i}$, such that the 
subspaces ${\cal G}_{\pm l_{\pm i}}$ are nontrivial. Restrict also the 
connection components $\omega_{\pm i} = \sum_{m = 0}^{\pm l_{\pm
i}} \omega_{\pm i, m}$ by the condition $\omega_{\pm i, \pm l_{\pm i}} = 
\zeta_\pm c_{\pm i}\zeta_\pm^{-1}$, where $\zeta_\pm$ are some mappings 
$M \to G_0$ with $G_0$ being the Lie group corresponding to ${\cal G}_0$, 
and $c_{\pm i}$ are some fixed 
elements of the subspaces ${\cal G}_{\pm l_{\pm i}}$ satisfying the relations 
$[c_{\pm i}, c_{\pm j}] = 0$. Then one can prove the following statement.

{\it There exists a local $G_0$--gauge transformation that brings a connection 
satisfying the above given conditions to the connection $\omega$ with the 
components} 
\begin{eqnarray*}
&\omega_{+i} = \gamma^{-1} \left(\sum_{m = 1}^{l_{+i} -1} \upsilon_{+i, m} +
c_{+i} \right) \gamma,& \\
&\omega_{-i} = \gamma^{-1} \partial_{-i} \gamma + \sum_{m =
-1}^{-l_{-i} + 1} \upsilon_{-i,m} + c_{-i},&  
\end{eqnarray*}
{\it where $\gamma$ is some mapping from $M$ to $G_0$, and $\upsilon_{\pm i, 
m}$ are mappings taking values in ${\cal G}_{\pm m}$.}

The equations for the mappings $\gamma$ and $\upsilon_{\pm i,m}$
which follow from the flatness condition we call multidimensional
Toda type systems, and the corresponding functions parametrising the
mappings $\gamma$ and $\upsilon_{\pm i,m}$ --- Toda and matter type
fields, respectively. In the proof we use the so called modified
Gauss decompositions\cite{RSa94} which allow to overcome the main
disadvantage of any standard Gauss decomposition that not any element
of $G$ possesses this decomposition. Namely, if an element $a \in G$
does not admit the Gauss decomposition of some form, then subjecting
$a$ to a left shift in $G$ we can get an element having this
decomposition.

Let us give examples of multidimensional Toda type equations, namely
those corresponding to the cases $l_- = l_+ = l = 1, 2$.  For $l= 1$
one has
\begin{eqnarray}
&[c_{\pm i}, \gamma^{\pm 1}\partial_{\pm j}\gamma^{\mp 1}]
- [c_{\pm j}, \gamma^{\pm 1}\partial_{\pm i}\gamma^{\mp 1}]=0,& \label{1}\\
&\partial_{+j} (\gamma^{-1} \partial_{-i} \gamma) = [c_{-i},
\gamma^{-1}c_{+j} \gamma ].& \label{2} 
\end{eqnarray}
For $l=2$ with the renotation $\upsilon_{\pm i, \pm 1}\equiv\upsilon_{\pm i}$
one has
\begin{eqnarray*}
&[c_{\pm i}, \upsilon_{\pm j}] = [c_{\pm j}, \upsilon_{\pm i}],& \\
&\partial_{\pm i}\upsilon_{\pm j}\pm [\gamma^{\pm 1}
\partial_{\pm i}\gamma^{\mp 1}, \upsilon_{\pm j}]=
\partial_{\pm j}\upsilon_{\pm i}\pm [\gamma^{\pm 1}
\partial_{\pm j}\gamma^{\mp 1}, \upsilon_{\pm i}],& \\
&[c_{\pm i}, \gamma^{\pm 1}\partial_{\pm j}\gamma^{\mp 1}]-
 [c_{\pm j}, \gamma^{\pm 1}\partial_{\pm i}\gamma^{\mp 1}]
+[\upsilon_{\pm i}, \upsilon_{\pm j}]=0;& \\
&\partial_{\pm i}\upsilon_{\mp j}= [c_{\mp j}, \gamma^{\mp 1}
\upsilon_{\pm i}\gamma^{\pm 1}],& \\
&\partial_{+j} (\gamma^{-1} \partial_{-i} \gamma) = 
[\upsilon_{-i}, \gamma^{-1}\upsilon_{+j} \gamma ]+
[c_{-i}, \gamma^{-1}c_{+j} \gamma ].& 
\end{eqnarray*}

{\bf 3.} Describe briefly the procedure for obtaining the general solution of 
the multidimensional Toda type equations. We start with mappings 
$\gamma_\pm$ taking values in $G_0$ and mappings $\lambda_{\pm i,
m}$ with values in ${\cal G}_{\pm m}$, which satisfy the conditions 
$\partial_{\mp i}\gamma_\pm =0$, $\partial_{\mp i}\lambda_{\pm i,
m}=0$, and the integrability conditions of the equations 
\begin{equation}
\mu^{-1}_{\pm}\partial_{\pm i}\mu_{\pm}=\sum_{m=1}^{l_{\pm i}-1}
\lambda_{\pm i, m}+\gamma_{\pm}c_{\pm i}\gamma_{\pm}^{-1}, \label{3}
\end{equation}
where $\mu_\pm$ are mappings from $M$ to $G$, such that
$\partial_{\pm i}\mu_\mp =0$.  The solution of equations (\ref{3})
is determined by the initial conditions $\mu_\pm(p) = a_\pm$, where
$p$ is some fixed point of $M$ and $a_\pm$ are some fixed elements of
$G$. It is clear that the mappings $\mu_\pm$ take values in the
subsets $a_\pm N_\pm$ with $N_\pm$ being the Lie subgroups of $G$
corresponding to the Lie subalgebras ${\cal N}_\pm = \oplus_{m > 0}
{\cal G}_{\pm m}$.  Further, the Gauss decomposition for
$\mu_+^{-1}\mu_-$ of the form $\mu_+^{-1} \mu_- = \nu_-\eta
\nu_+^{-1}$ gives the mappings $\eta$ and $\nu_\pm$ which take values
in $G_0$ and $N_{\pm}$, respectively. Finally, using the formula
$\gamma =\gamma_+^{-1}\eta\gamma_-$ and the decompositions
\[
\sum_{m =1}^{l_{\pm i}}\upsilon_{\pm i,m}=\gamma_{\pm}^{-1}\eta^{\pm 1}
(\nu_{\pm}^{-1}\partial_{\pm i}\nu_{\pm})\eta^{\mp 1}\gamma_{\pm},\]
we obtain the mappings $\gamma$ and $\upsilon_{\pm i, m}$ which
satisfy the multidimensional Toda type equations. Any solution
can be obtained using this procedure. In general, different sets of mappings
$\gamma_\pm$, $\lambda_{\pm i, m}$, as well as different choices of
initial conditions 
for $\mu_{\pm}$, can give the same solution. Note that almost all solutions 
of the multidimensional Toda type equations can be obtained using the
mappings $\mu_\pm$ taking values in the subgroups $N_\pm$.

{\bf 4.} As an illustration of our general construction consider
a particular case of system (\ref{1}), (\ref{2}) corresponding
to the loop group ${\cal L}(GL(2m, {\bf C}))$. With an appropriate
{\bf Z}--gradation of the 
Lie algebra ${\cal L}({\cal GL}(2m, {\bf C}))$ one has
\[
c_{-i} = \left(\begin{array}{cc}
0 & \zeta^{-1} K_{-i} \\
L_{-i} & 0
\end{array} \right), \quad
c_{+i} = \left(\begin{array}{cc}
0 & K_{+i} \\
\zeta L_{+i} & 0
\end{array} \right), \quad
\gamma = \left( \begin{array}{cc}
\beta_1 & 0 \\
0 & \beta_2
\end{array} \right).
\]
Here each entry in the matrices is an $m\times m$ block, $\beta_{1,2}$
take values in $GL(m, {\bf C})$, $\zeta$ is the loop parameter.
By this example we show that subsystem (\ref{1}) gives two sets of
generalised wave equations over the coordinates $z^{+i}$ and
$z^{-i}$, respectively, while subsystem (\ref{2}) is a dynamical one
and provides a nontrivial mixing of the dependence on the whole set
of the coordinates.

If $L_{\pm i} = K_{\pm i}^t$, the system under consideration 
admits the reduction to the case when $\beta_{1,2}$ take values in
the complex orthogonal group $O(m, {\bf C})$, which leads to the equations
\begin{eqnarray}
&\partial_{-i} (\beta_1 K_{-j} \beta_2^{-1}) = 
\partial_{-j} (\beta_1 K_{-i}
\beta_2^{-1}),& \label{14} \\
&\partial_{+i} (\beta_1^{-1} K_{+j}\beta_2) = 
\partial_{+j} (\beta_1^{-1} K_{+i}
\beta_2),& \label{4}\\
&\partial_{+j} (\beta_{1}^{-1}\partial_{-i}\beta_{1})= 
K_{-i}\beta_{2}^{-1}K_{+j}^t \beta_{1}-
\beta_{1}^{-1} K_{+j}\beta_{2}K_{-i}^t,& \label{5}\\
&\partial_{+j} (\beta_{2}^{-1}\partial_{-i}\beta_{2})= 
K_{-i}^t \beta_{1}^{-1}K_{+j}\beta_{2}-
\beta_{2}^{-1} K_{+j}^t \beta_{1}K_{-i}. \label{15}
\end{eqnarray}
For $K_{\pm i}^t = K_{\pm i}$ system
(\ref{14})--(\ref{15}) admits further reduction to the case $\beta_1
= \beta_2 = \beta$, and one ends up with the
Cecotti--Vafa equations\cite{CVa91},
\begin{eqnarray}
&\partial_{-i} (\beta K_{- j} \beta^{-1}) = 
\partial_{- j} (\beta K_{- i}\beta^{-1}),& \label{11} \\
&\partial_{+ i} (\beta^{- 1}K_{+ j}\beta) = 
\partial_{+j} (\beta^{- 1}K_{+i}\beta),& \label{16} \\
&\partial_{+j} (\beta^{-1} \partial_{-i} \beta) = 
[K_{-i}, \beta^{-1} K_{+j} \beta ].& \label{12}
\end{eqnarray}
Equations (\ref{11}), (\ref{16}) with $(K_{\pm
i})_{kl}=\delta_{ik}\delta_{il}$ are connected with some well known
equations in differential geometry\cite{Dub93}. With the same choice
of $K_{\pm i}$, a similar connection takes place for more general
equations (\ref{14}), (\ref{4}). Namely, introduce the notation 
\[
b_{ij} = \frac{1}{(\beta_1)_{mi}} \partial_{+i} (\beta_1)_{mj}, \qquad i \ne j.
\]
It is not
difficult to show that equations (\ref{4}) are equivalent to the
relations
\begin{equation}
\beta_1^{-1}\partial_i\beta_1= K_i b - b^t K_i,\qquad
\beta_2^{-1}\partial_i\beta_2= K_i b^t - b K_i. \label{13}
\end{equation}
Here and in what follows for the sake of brevity we omit signs $\pm$
in the indices. The functions $b_{ij}$ defined in such a way satisfy 
the equations
\begin{eqnarray}
& \partial_ib_{ji}+\partial_jb_{ij}+\sum_{k\neq i,j}
b_{ik}b_{jk}=0,\; i\neq j;&\label{6}\\
&\partial_kb_{ji}=b_{jk}b_{ki},\; 
i\neq j\neq k;&\label{7}\\
&\partial_ib_{ij}+\partial_jb_{ji}+\sum_{k\neq i,j}
b_{ki}b_{kj}=0,\; i\neq j,&\label{8}
\end{eqnarray}
which are nothing but the zero curvature condition for the connections
with components $\beta^{-1}_{1,2} \partial_i \beta_{1,2}$ given by
(\ref{13}). From the other hand, if we have a solution of equations
(\ref{6})--(\ref{8}), then integrating equations (\ref{13}) we come to
the mappings $\beta_{1,2}$ which satisfy equations (\ref{4}). 
When $\beta_1 = \beta_2$, it appears that
the Egorov property, $b_{ij} = b_{ji}$, is valid, and equations
(\ref{8}) are reduced to 
\[
\sum_{k=1}^m\partial_kb_{ij}=0.
\]

Equations (\ref{6})--(\ref{8}) are related to some classical problems
of differential geometry, moreover, they are completely
integrable\cite{Dar10}. 

The system arising from equations (\ref{5}), (\ref{15}) with the
specialisation chosen above, sometimes is called a multidimensional
generalisation of the sine--Gor\-don equation, while those from
(\ref{8}) --- generalised wave equations. In particular, for the
simplest case with $m=2$, the integration problem of these equations
can be reduced to the solution of two independent sine--Gordon equations.


\section*{References}
\begin{thebibliography}{99}

\bibitem{LSa92} A. N. Leznov and M. V. Saveliev, {\em Group--Theoretical 
Methods for Integration of Nonlinear Dynamical Systems} (Birkhauser-Verlag, 
Basel, 1992).

\bibitem{RSa94} A. Razumov and M. Saveliev, \Journal{Comm. 
Anal. Geom.}{2}{461}{1994}.

\bibitem{GSa95} J.-L. Gervais and M. V. Saveliev, \Journal{Nucl. Phys.}
{B453}{449}{1995}.

\bibitem{FGGS95} L.A. Ferreira, J.-L. Gervais, J. S. Guillen and M. V. 
Saveliev, {\em Affine  Toda systems coupled to matter fields}, 
hep--th/9512105 (to appear in Nucl. Phys. B).  

\bibitem{CVa91} S. Cecotti and C. Vafa, \Journal{Nucl. Phys.}{B367}{359}{1991}.

\bibitem{Dub93} B. Dubrovin, \Journal{Comm. Math. Phys.}{152}{539}{1993}.

\bibitem{GMa93} J.-L. Gervais and Y. Matsuo, \Journal{Comm. Math. Phys.}{152}
{317}{1993}.

\bibitem{Dar10}
G. Darboux, {\em Lecons sur les syst\`emes orthogonaux et les coordonn\'ees
curvilignes} (Gauthier--Villars, Paris, 1910). 

\end{thebibliography}
\end{document}